\documentclass[a4paper,french, submission]{rnti}
\usepackage[T1]{fontenc}
\usepackage[latin1]{inputenc}

\usepackage{url}

\usepackage{graphicx}

\usepackage[bookmarks=true, bookmarksnumbered=true, bookmarksopen=true,
		unicode=true, colorlinks=true,
		pagebackref=true,
		linkcolor=blue,citecolor=blue,filecolor=blue,urlcolor=blue
		]{hyperref}

\titrecourt{Rôles communautaires dans les réseaux orientés}

\nomcourt{N. Dugué et al.}

\titre{Identification de rôles communautaires dans des réseaux orientés appliquée à Twitter}

\auteur{Nicolas Dugué\affil{1},
        Vincent Labatut\affil{2},
        Anthony Perez\affil{1}}

\affiliation{
    \affil{1}Universit\'e d'Orl\'eans, ENSI de Bourges, LIFO EA 4022, F-45067 Orl\'eans, France\\
    \affil{2}Galatasaray University, Computer Science Department, \c C{\i}ra\u{g}an cad. n°36, Ortak\"oy 34357, \.{I}stanbul, Turquie\\
    \affil{3}Encore une autre adresse\\
          encore-une-autre-adresse@email
 }

\resume{%
La notion de structure de communautés est particulièrement utile pour étudier les réseaux complexes, car elle amène un niveau d'analyse intermédiaire, par opposition aux plus classiques niveaux local (voisinage des n\oe{}uds) et global (réseau entier). Le concept de rôle communautaire a été dérivé sur cette base, afin de décrire le positionnement d'un n\oe{}ud en fonction de sa connectivité communautaire. Cependant, les approches existantes sont restreintes aux réseaux non-orientés, elles utilisent des mesures topologiques ne considérant pas tous les aspects de la connectivité communautaire, et des méthodes d'identification des rôles non-généralisables à tous les réseaux. Nous proposons de résoudre ces problèmes en généralisant et étendant les mesures existantes, et en utilisant une méthode non-supervisée pour déterminer les rôles. Nous illustrons l'intérêt de notre méthode en l'appliquant à l'analyse du réseau de Twitter. Nous montrons que nos modifications permettent de mettre en évidence les rôles spécifiques d'utilisateurs particuliers du réseau, nommés capitalistes sociaux.
}

\summary{%
The notion of community structure is particularly useful when analyzing complex networks, because it provides an intermediate level, compared to the more classic global (whole network) and local (node neighborhood) approaches. The concept of community role of a node was derived from this base, in order to describe the position of a node in a network depending on its connectivity at the community level. However, the existing approaches are restricted to undirected networks, use topological measures which do not consider all aspects of community-related connectivity, and their role identification methods are not generalizable to all networks. We tackle these limitations by generalizing and extending the measures, and using an unsupervised approach to determine the roles. We then illustrate the applicability of our method by analyzing a Twitter network. We show how our modifications allow discovering the fact some particular users called \textit{social capitalists} occupy very specific roles in this system.
}

\begin{document}
\section{Introduction}
Les réseaux complexes sont des graphes modélisant des systèmes réels. Leurs propriétés topologiques ont récemment fait l'objet de nombreux travaux, dont un certain nombre s'est concentré sur leur structure de communautés \citep{Fortunato2010}. Dans sa forme la plus simple, cette structure est une partition de l'ensemble des n\oe{}uds, dont les parties (communautés) sont des groupes de n\oe{}uds densément interconnectés. La notion de communauté est particulièrement intéressante, car elle permet l'étude du réseau à un niveau intermédiaire, par comparaison avec les plus classiques niveaux local (voisinage du n\oe{}ud) et global (réseau entier). Le concept de r\^ole communautaire illustre bien cette caractéristique : il décrit la position d'un n\oe{}ud dans sa communauté. Il a été initialement introduit par~\cite{Guimera2005}, puis indépendamment par \cite{Scripps2007}. En se basant sur une estimation de la structure de communautés, ces auteurs caractérisent le positionnement communautaire de chaque n\oe{}ud au moyen de plusieurs mesures topologiques \textit{ad hoc}. Les n\oe{}uds sont ensuite catégorisés au moyen de seuils prédéfinis pour chaque mesure. 

Ces approches peuvent être critiquées sur trois points. Premièrement, elles sont définies seulement pour des réseaux non-orientés. Pourtant, de nombreux systèmes contiennent des relations asymétriques, et ne pas en tenir compte constitue une perte significative d'information. Deuxièmement, les mesures utilisées ne prennent pas en compte tous les aspects de la connectivité communautaire d'un n\oe{}ud. Troisièmement, rien ne garantit que les seuils fixés empiriquement pour définir les rôles soient pertinents pour d'autres données. Dans ce travail, nous proposons des solutions à ces trois problèmes. Pour le premier, nous adaptons les mesures de Guimer\`a \& Amaral aux réseaux orientés. Pour le deuxième, nous définissons des mesures supplémentaires distinguant trois aspects de la connectivité communautaire : diversité des communautés, hétérogénéité de la distribution des liens, et intensité de la connexion. Pour le troisième, nous proposons une méthode non-supervisée de définition des rôles, utilisant les mesures proposées. 

Afin d'illustrer l'intérêt de notre méthode, nous l'appliquons à l'étude du rôle communautaire d'un type particulier d'utilisateur de Twitter, appelé \textit{capitaliste social}. Le principe du capitalisme social est d'essayer d'obtenir un maximum de visibilit\'e en utilisant diverses techniques. Cette notion a été mise en évidence dans un travail sur les comptes spammers de Twitter par \cite{GVK+12}. Sur Twitter, deux principes relativement simples sont principalement utilisés pour accro\^itre le nombre de followers et donc la visibilit\'e. \textit{Follow Me, I Follow You} (FMIFY) : le capitaliste promet aux utilisateurs qui le suivent de les suivre en retour. \textit{I Follow You, Follow Me} (IFYFM) : le capitaliste suit un maximum d'utilisateurs, en esp\'erant \^etre suivi en retour. De tels utilisateurs peuvent \^etre n\'efastes pour l'\'equilibre du r\'eseau social, dans la mesure o\`u leurs comptes gagnent en visibilit\'e et leurs tweets sont bien class\'es par les moteurs de recherche du r\'eseau, mais souvent sans r\'eelle raison de contenu. L'amélioration de la qualité du service passe donc par une bonne compréhension de leur positionnement dans le réseau, et donc dans les communautés. On peut en effet se demander si ces utilisateurs sont ancr\'es dans leur communaut\'e, \'etroitement li\'es aux autres utilisateurs, ou s'ils sont au contraire isol\'es. Une autre question est de savoir s'ils sont li\'es aux autres communaut\'es, et, si oui, avec quelle intensit\'e.

Dans la section suivante, nous décrivons l'approche originale de Guimer\`a \& Amaral. Nous mettons ensuite en évidence ses limitations, et proposons notre propre approche en section \ref{sec:propose}. Dans la section \ref{sec:resultats}, nous présentons les rôles obtenus sur le réseau Twitter et discutons du positionnement des capitalistes sociaux. Enfin, nous concluons en indiquant les perspectives ouvertes par ce travail.

\section{Approche originale}
Nous avons décidé de construire notre méthode à partir de celle de \cite{Guimera2005}, non seulement parce qu'elle est plus répandue que celle de \cite{Scripps2007}, mais aussi parce qu'elle s'appuie plus fortement sur la structure de communauté. Pour caractériser les rôles des n\oe{}uds, Guimer\`a \& Amaral définissent d'abord deux mesures complémentaires, qui leur permettent de placer chaque n\oe{}ud dans un espace bidimensionnel. Puis, ils proposent plusieurs seuils pour discrétiser cet espace, chaque zone ainsi définie correspondant à un rôle particulier. Dans cette section, nous décrivons d'abord les mesures, puis la méthode qu'ils utilisent pour identifier les rôles.

%%\subsection{Degré intra-module et coefficient de participation}
La première mesure, nommée \textit{degré intra-module} (\textit{within-module degree} en anglais) traite de la connectivité interne du n\oe{}ud, i.e. des liens avec sa propre communauté. Elle est basée sur la notion de $z$-score. Comme celle-ci sera réutilisée plus loin, nous la définissons ici de façon générique. Pour une fonction nodale quelconque $f(u)$, permettant d'associer une valeur numérique à un n\oe{}ud $u$, le $z$-score $Z_f(u)$ par rapport à la communauté de $u$ est :
\begin{equation}
\label{f:zscore}
Z_f(u) = \frac{f(u) - \mu_i(f)} {\sigma_i(f)} 
\mbox{, avec } u \in C_i
\end{equation}
où $C_i$ représente une communauté, et $\mu_i(f)$ et $\sigma_i(f)$ dénotent respectivement la moyenne et l'écart-type de $f$ sur les n\oe{}uds appartenant à la communauté $C_i$. Le degré intra-module de Guimerà et Amaral, noté $z(u)$, correspond au $z$-score du degré interne, calculé pour la communauté du n\oe{}ud considéré. On l'obtient donc en substituant le degré interne $d_{int}$ à $f$ dans l'équation (\ref{f:zscore}). Le degré intra-module évalue la connectivité d'un n\oe{}ud à sa communauté relativement à celle des autres n\oe{}uds de sa communauté. La seconde mesure, appelée \textit{coefficient de participation}, traite de la connectivité externe du n\oe{}ud, i.e. relative à toutes les communautés auquel il est lié. Elle est définie de la manière suivante :
\begin{equation}
\label{eq:p}
P(u) = 1 - \sum_i{\left(\frac{d_i(u)}{d(u)}\right)^2}
\end{equation}
où $d_i(u)$ représente le nombre de liens que $u$ possède vers des n\oe{}uds de la communauté $C_i$. Notons que dans le cas où $C_i$ est la communauté de $u$, alors on a $d_i(u) = d_{int}(u)$. Le coefficient de participation représente combien les connexions d'un n\oe{}ud sont diversifiées, en termes de communauté externes. Une valeur proche de $1$ signifie que le n\oe{}ud est connecté de façon uniforme à un grand nombre de communautés différentes. Au contraire, une valeur de $0$ ne peut être atteinte que si le n\oe{}ud n'est connecté qu'à une seule communauté (vraisemblablement la sienne).

\cite{Guimera2005} proposent de caractériser le rôle d'un n\oe{}ud dans un réseau en se basant sur ces deux mesures. Pour ce faire, ils définissent sept rôles différents en discrétisant l'espace à deux dimensions formé par $z$ et $P$. Un premier seuil défini sur le degré intra-module $z$ permet de distinguer ce que les auteurs appellent les \textit{pivots communautaires} ($z\geq2.5$) des autres n\oe{}uds ($z<2.5$). Ces pivots (\textit{hubs} en anglais) sont considérés comme fortement intégrés à leur communauté, par rapport au reste des n\oe{}uds de cette même communauté. Ces deux catégories (pivot et non-pivot) sont subdivisées au moyen d'une série de seuils définis sur le coefficient de participation $P$. En considérant les n\oe{}uds par participation croissante, Guimerà et Amaral les qualifient de \textit{provinciaux} ou \textit{(ultra-)périphériques}, \textit{connecteurs} et \textit{orphelins}. Les deux premiers rôles sont essentiellement connectés à leur communauté, les troisièmes, bien qu'eux aussi potentiellement bien connectés à leur propre communauté, sont également largement liés à d'autres communautés, et les derniers sont connectés à un grand nombre de communautés.

%TODO VL je ne sais pas où le texte ci-dessous devait aller, mais ce n'était surement pas ici en tout cas !
%Dans la mesure o\`u le r\'eseau repr\'esentant Twitter est orient\'e, nous adoptons les mesures pr\'esent\'ees pr\'ec\'edemment afin d'\'eviter toute perte d'information li\'ee aux informations contenues dans l'orientation des relations. 

\section{Méthode proposée}
\label{sec:propose}
Dans cette section, nous décrivons les trois modifications que nous proposons pour résoudre les limitations précédemment mentionnées de l'approche de Guimer\`a \& Amaral. Nous expliquons tout d'abord comment étendre les mesures pour le cas d'un réseau orienté, puis proposons des mesures supplémentaires permettant de mieux évaluer la connectivité externe des n\oe{}uds, et enfin une méthode non-supervisée pour déterminer les rôles nodaux.

\subsection{Orientation des liens}
\label{subsec:oriente}
Il est souvent assez simple de g\'en\'eraliser des mesures d\'efinies sur des graphes non-orient\'es vers des graphes orient\'es. En effet, la sch\'ema classique consiste \`a distinguer les liens \emph{entrants} des liens \emph{sortants}. Dans notre cas, cela consister à utiliser $4$ mesures au lieu de $2$: degr\'es intra-module entrant et sortant, ainsi que coefficients de participation entrant et sortant.

Nous notons $d^{in}$ le degr\'e entrant d'un n\oe{}ud, i.e. le nombre de liens entrants connect\'es \`a ce n\oe{}ud. Nous pouvons ainsi d\'efinir le \emph{degr\'e entrant interne} d'un n\oe{}ud, not\'e $d^{in}_{int}$ et repr\'esentant le nombre de liens entrants que le n\oe{}ud poss\`ede \`a l'int\'erieur de sa communaut\'e. En calculant le $z$-score de cette valeur, nous obtenons ainsi le \emph{degr\'e intra-module entrant}, not\'e $z^{in}$. De mani\`ere similaire, nous d\'efinissons $d_i^{in}$ comme le \emph{degr\'e communautaire entrant}, \`a savoir le nombre de liens entrants qu'un n\oe{}ud a avec les n\oe{}uds de la communaut\'e $C_i$. Cela nous permet de d\'efinir le \emph{coefficient de participation entrant}, not\'e $P^{in}$, en rempla\c cant $d$ par $d^{in}$ et $d_i$ par $d_i^{in}$ dans l'\'equation (\ref{eq:p}). Le \emph{degr\'e intra-module sortant} $z^{out}$ et le \emph{coefficient de participation sortant} $P^{out}$ sont obtenus de façon symétrique, en utilisant les contreparties sortantes des degrés entrants : $d^{out}$, $d^{out}_{int}$ et $d_i^{out}$.

Remarquons ici qu'il ne serait pas pertinent de calculer de telles mesures en se basant sur un ensemble de communaut\'es d\'etect\'e par un algorithme ne tenant pas compte de l'orientation des liens. Il est donc n\'ecessaire d'utiliser un algorithme prenant en compte cette information. Un autre point important est la d\'efinition des r\^oles. Il n'y a a priori aucune raison pour que les seuils d\'efinis par~\cite{Guimera2005} soient toujours valables pour les versions orientées des mesures. Nous verrons cependant que la m\'ethode d'identification des r\^oles non supervis\'ee que nous proposons dans la Section~\ref{subsec:roles} permet de r\'esoudre ce probl\`eme.

\subsection{Aspects de la connectivité externe}
\label{subsec:nosmesures}
\begin{figure}[h]
	\begin{minipage}{0.32 \linewidth}
		\centerline{\includegraphics[scale=0.10]{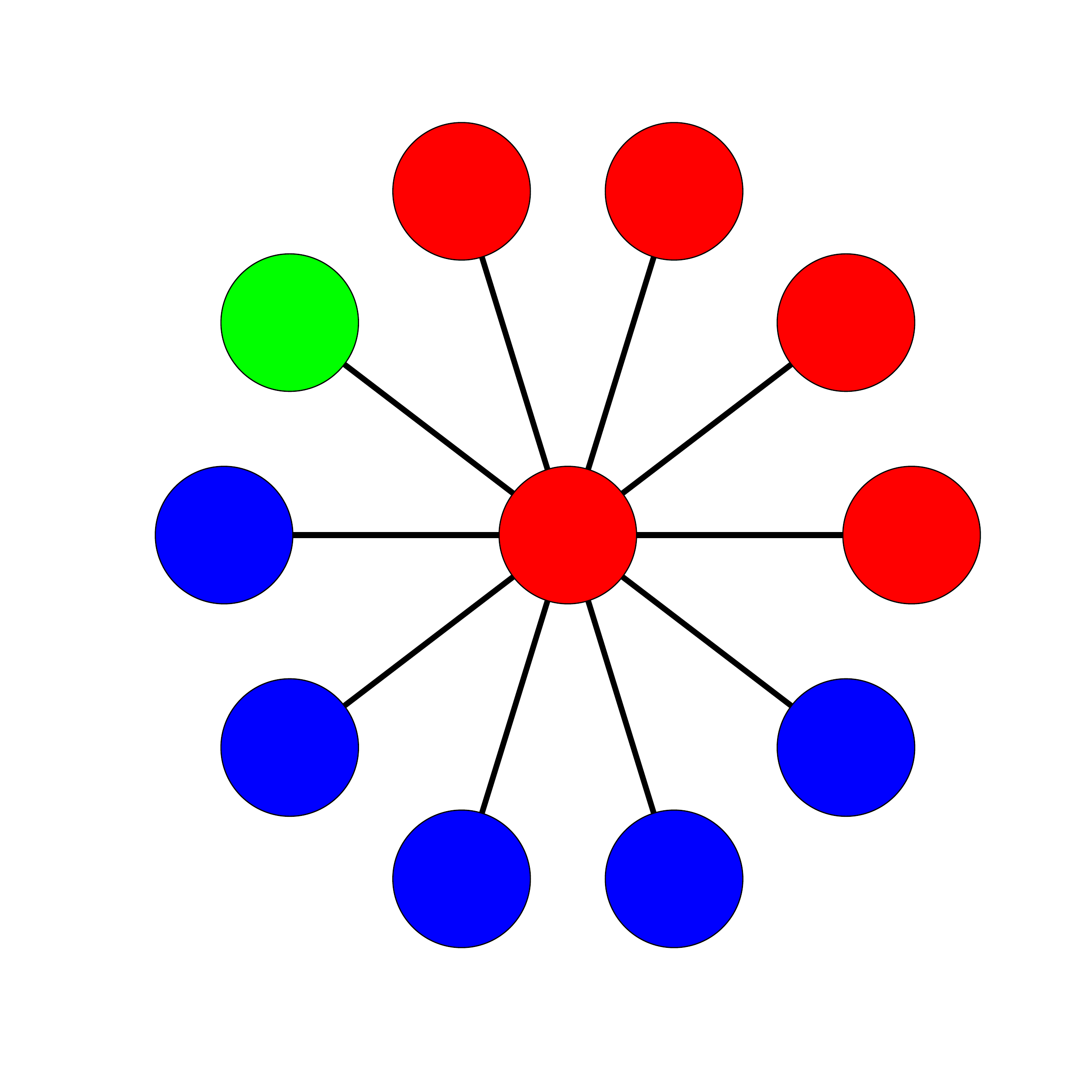}}
	\end{minipage}
	\begin{minipage}{0.32 \linewidth}
		\centerline{\includegraphics[scale=0.10]{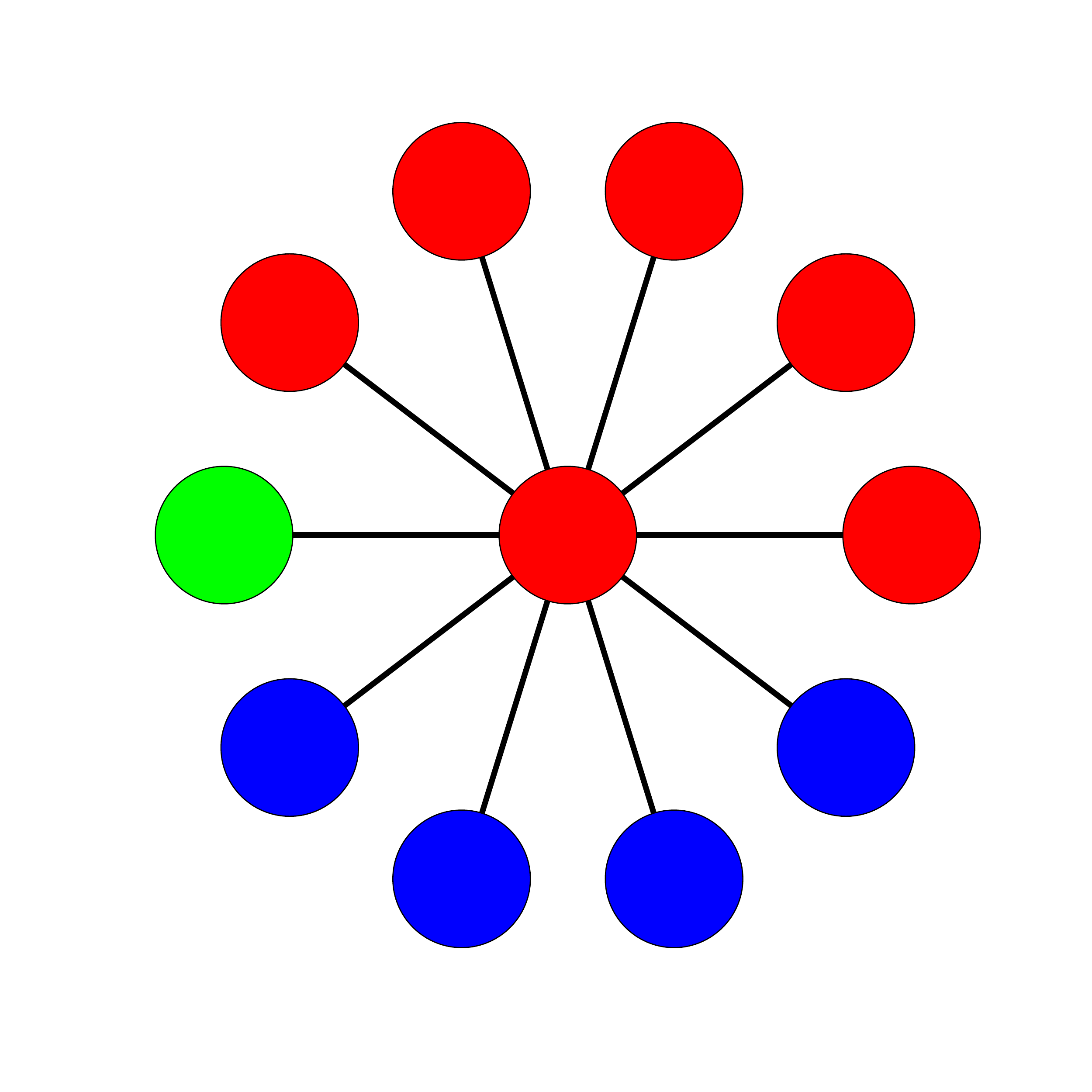}}
	\end{minipage}
	\begin{minipage}{0.32 \linewidth}
		\centerline{\includegraphics[scale=0.10]{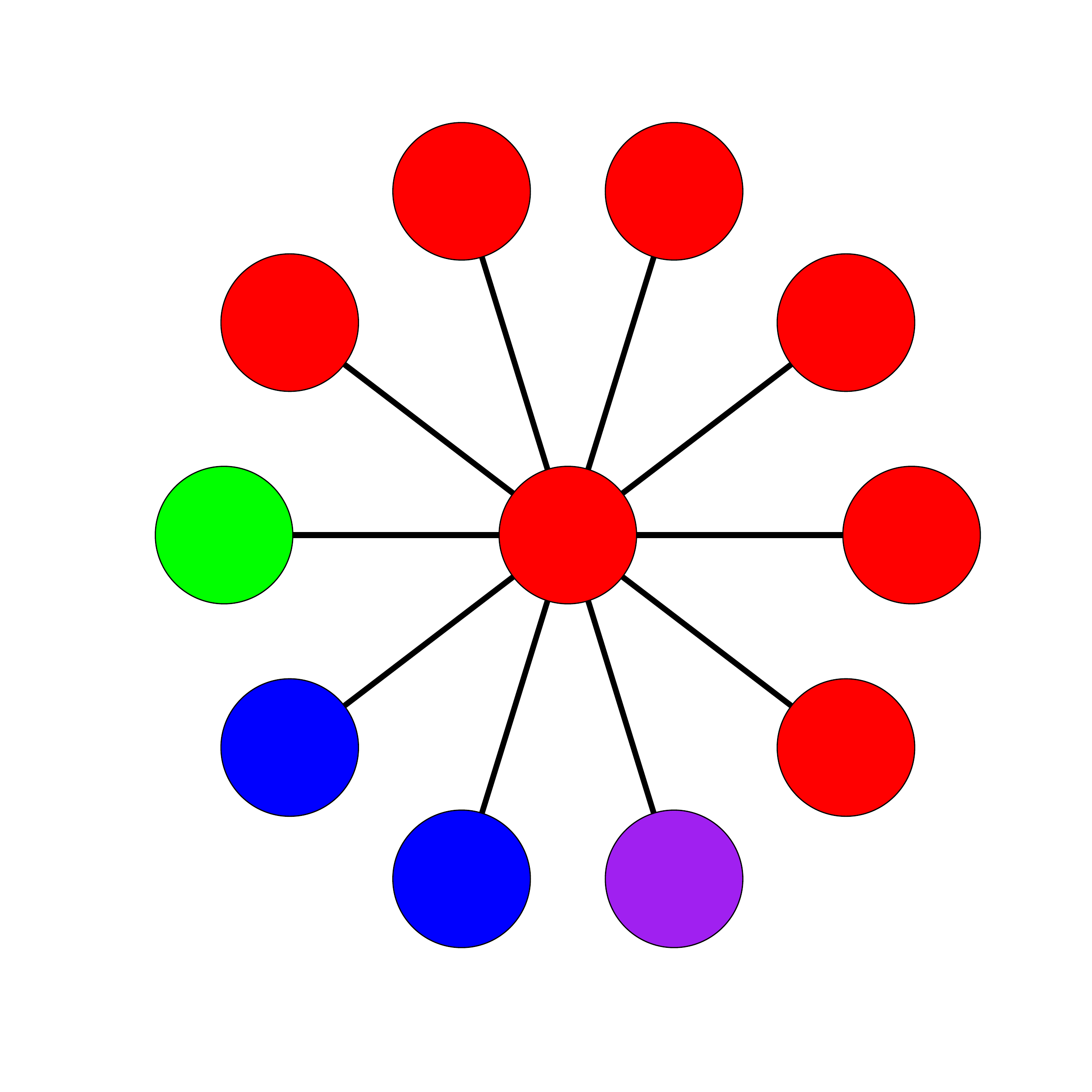}}
	\end{minipage}
	
		\caption{Chaque couleur repr\'esente une communaut\'e. Dans chaque cas, le coefficient de participation du n\oe{}ud central est $0,58$.
	\label{fig:participation}}
\end{figure}

Le coefficient de participation se concentre sur un aspect de la connectivité externe d'un n\oe{}ud : l'\textit{hétérogénéité} de la distribution de ses liens, relativement aux communautés auxquelles il est connecté. Mais il est possible de caractériser cette connectivité de deux autres manières. Premièrement, on peut considérer sa \textit{diversité}, c'est à dire le nombre de communautés concernées. Deuxièmement, il est possible de s'intéresser à son \textit{intensité}, i.e. au nombre de liens concernés. Comme le montre la Figure~\ref{fig:participation}, ces deux aspects ne sont pas pris en compte dans $P$. En effet, bien que la connectivit\'e externe du n\oe{}ud central soit diff\'erente sur chacune des trois figures, le coefficient de participation reste le m\^eme. Pour pallier cette limitation, nous proposons deux nouvelles mesures permettant de quantifier la diversité et l'intensité. De plus, afin d'obtenir un ensemble cohérent de mesures, nous révisons également $P$.

\paragraph{Diversité.} Notre mesure de \textit{diversité}, notée $D(u)$, \'evalue le nombre de communaut\'es diff\'erentes auxquelles le n\oe{}ud $u$ est connect\'e %quantifie combien le n\oe{}ud $u$ est connecté à un grand nombre de communautés différentes
, indépendamment de la densité de ces connexions. Soit $\epsilon(u)$ le nombre de communautés, autres que la sienne, auxquelles le n\oe{}ud $u$ est connecté. Alors la diversité est définie comme le $z$-score d'$\epsilon$ relativement à la communauté de $u$. C'est à dire qu'on l'obtient en substituant $\epsilon$ à $f$ dans (\ref{f:zscore}).

\paragraph{Intensité externe.} L'\textit{intensité externe} $I_{ext}(u)$ mesure la force de la connexion de $u$ à des communautés externes, en termes de nombre de liens, et relativement aux autres n\oe{}uds de sa communauté. Soit $d_{ext}(u)$ le degré externe de $u$, correspondant au nombre de liens que $u$ possède avec des n\oe{}uds n'appartenant pas à sa communauté. Remarquons qu'on a alors $d=d_{int}+d_{ext}$. Nous définissons l'intensité externe comme le $z$-score du degré externe, c'est à dire qu'on l'obtient en substituant $d_{ext}$ à $f$ dans (\ref{f:zscore}).

\paragraph{Hétérogénéité.} L'\textit{hétérogénéité} $H(u)$ quantifie combien le nombre de connexions externes du n\oe{}ud $u$ varie d'une communauté à l'autre. Nous utilisons pour cela l'écart-type du nombre de liens externes que le n\oe{}ud possède par communauté, que nous notons $\lambda(u)$. L'hétérogénéité est alors le $z$-score de $\lambda$, relativement à la communauté de \textit{u}, et on l'obtient donc en substituant $\lambda$ à $f$ dans (\ref{f:zscore}). Cette mesure a une signification très proche de celle du coefficient de participation $P$ de Guimerà et Amaral. Elle diffère en ce qu'elle est exprimée relativement à la communauté de $u$, et que les liens internes à cette même communauté sont exclus du calcul.

\paragraph{Intensité interne.} Pour représenter la connectivité interne du n\oe{}ud, nous conservons la mesure $z$ de Guimerà et Amaral. En effet, celle-ci est construite sur la base du $z$-score, et est donc cohérente avec les autres mesures définies pour décrire la connectivité externe. De plus, il n'est pas nécessaire de lui adjoindre d'autres mesures, car les notions d'hétérogénéité et de diversité n'ont pas de sens ici (puisqu'on considère seulement une seule communauté). Cependant, en raison de sa symétrie avec notre intensité externe, nous désignons $z$ sous le nom d'\textit{intensité interne}, et la notons $I_{int}(u)$. \\

Pour chacune des 4 mesures présentées, nous utilisons deux variantes, l'une considère les liens entrants, l'autre les liens sortants. 

\subsection{Identification non-supervisée des r\^oles}
\label{subsec:roles}
Notre derni\`ere modification de l'approche de ~\cite{Guimera2005} concerne la mani\`ere dont les r\^oles sont définis. \cite{Guimera2005} supposent l'existence de rôles universels, présents dans tous les systèmes. %Cependant, lors de la sélection de seuils leur permettant de définir ces rôles, ils supposent également que ces seuils devraient être valables indépendamment du réseau considéré.
Ils supposent notamment que les seuils établis de façons empiriques pour définir les r\^oles sont indépendants des jeux de données utilisés. Cette dernière partie est sujette à discussion, dans la mesure où, parmi leurs deux mesures, seules $P$ est normalisée sur un intervalle fixé. En effet, il n'y a aucune limitation pour $z$, et il n'y a donc aucune garantie que le seuil défini originellement pour cette mesure reste cohérent pour d'autres réseaux. Si nous considérons les mesures présentées dans la section~\ref{subsec:nosmesures}, cet argument est d'autant plus fort que \emph{toutes} nos mesures sont définies comme des $z$-scores. De plus, le fait que nous proposions $8$ mesures fait croître le nombre de seuils nécessaires de manière significative, et rend la sélection de seuil originelle impossible à utiliser.

Afin de contourner ces problèmes, nous proposons d'appliquer une méthode automatique de classification non supervisée. Dans un premier temps, nous calculons l'ensemble des mesures sur les données considérées. Ensuite, nous appliquons une analyse de regroupement. Chaque groupe ainsi identifié correspond a un rôle communautaire. Cette méthode présente l'avantage de ne pas être affectée par le nombre de mesures utilisé, et revient à ajuster les seuils pour le système considéré.

\section{Résultats}
\label{sec:resultats}

Le réseau sur lequel nous avons travaillé a été collecté en 2009 par \cite{CHBG10}. Il comporte un peu moins de $55$ millions de n\oe{}uds représentant les utilisateurs de Twitter et près de $2$ milliards d'arcs orientés qui matérialisent les abonnements entre utilisateurs, à savoir les liens de "follow". La très grande taille de ces données a influencé le choix de nos outils d'analyse. La détection de communautés a été réalisée au moyen de l'algorithme de Louvain~\citep{Blondel2008}, très efficace pour le traitement de grands réseaux. Nous avons repris le code mis à disposition par ses auteurs et l'avons adapté à la modularité orientée décrite par \cite{Newman2008}. L'analyse de regroupement a alors été menée au moyen d'une implémentation libre et distribuée de l'algorithme des $k$-moyennes \citep{Liao2009}. %En effet, les méthodes non-distribuées, basées sur le calcul d'une unique matrice de distance, se sont révélées impossible à appliquer en raison de la quantité de mémoire nécessaire à la représentation de la matrice. 
Nous avons appliqué cet algorithme pour des valeurs de $k$ allant de $2$ à $15$, et avons sélectionné la meilleure partition d'après l'indice de \cite{Davies1979}. %Les scripts de pré- et post-traitement relatifs à l'analyse de regroupement on été implémentés en langage R. 
L'ensemble de notre code source est disponible à l'adresse 
\url{https://github.com/CompNet/Orleans}. %TODO VL pour le package hyperref
%\texttt{https://github.com/CompNet/Orleans}.

Pour valider les résultats obtenus, nous étudions la position des capitalistes sociaux dans les r\^oles détectés. Ceux-ci sont identifiés en utilisant la méthode proposée par \cite{DUGUE2013}, basée sur l'utilisation de deux mesures topologiques spécifiques. Pour faciliter l'interprétation des résultats, à l'instar de Dugu\'e et Perez, nous distinguons différentes catégories de capitalistes sociaux en fonction de deux de leurs caractéristiques topologiques. La première est le \textit{ratio}. Il s'agit du nombre de followees divis\'e par le nombre de followers. Ce critère permet de distinguer ceux qui appliquent la méthode FMIFY (ratio inférieur à 1) de ceux utilisant IFYFM (ratio supérieur à 1). La seconde est le degré entrant: nous séparons ceux de faible degré (entre $500$ et $10000$) et ceux de degré élevé (supérieur à $10000$).

\subsection{Approche originale}
Nous avons tout d'abord appliqué l'approche originale (non-orientée) de Guimer\`a \& Amaral sur nos données. Les valeurs de \textit{z} obtenues sont bien supérieures à celles observées dans~\citep{Guimera2005}. Le seuil défini pour \textit{z} n'est ainsi plus utilisable pour l'identification des r\^oles. Nous avons donc procédé à une analyse de regroupement qui identifie $2$ r\^oles, contenant chacun trop de n\oe{}uds pour obtenir une information pertinente sur la connectivité des n\oe{}uds du réseau relativement à la structure de communautés. La perte d'information due au fait que la méthode originale ne tienne pas compte de l'orientation des liens peut expliquer ces résultats.

Nous avons ensuite appliqué l'approche originale adaptée aux graphes orientés, telle que décrite dans la Section~\ref{subsec:oriente}. L'analyse de regroupement a identifié $6$ r\^oles : un groupe de n\oe{}uds pivots (n\oe{}uds mieux connectés que les autres à leur communauté), et $5$ groupes de n\oe{}uds non-pivots. Les n\oe{}uds non-pivots sont séparés selon la distribution de leurs liens externes en utilisant les coefficients de participation. On retrouve ainsi des non-pivots périphériques et ultra-périphériques, considérés comme peu connectés aux communautés externes (de faibles $P^{in}$ ou $P^{out}$), et des non-pivot orphelins connectés de façon homogène avec les communautés externes ($P^{in}$ ou $P^{out}$ élevés). La diversité des r\^oles obtenus montre l'intér\^et des mesures orientées par rapport à l'approche non-orientée, qui avait considéré comme équivalents plusieurs de ces groupes. Néanmoins, lorsque l'on regarde le positionnement des capitalistes sociaux au sein de ces groupes, certaines incohérences apparaissent. Une large majorité des capitalistes sociaux de degré élevé est ainsi classée comme étant non-pivots périphériques ou ultra-périphériques. Ces n\oe{}uds ayant un degré entrant supérieur à $10000$, et pour certains un degré sortant supérieur, cela semble surprenant. En effet, si ces n\oe{}uds ne sont pas pivots, donc peu connectés, ils devraient néanmoins \^etre connectés avec l'extérieur. Cela vient du fait que la participation évalue l'\textit{hétérogénéité} des connexions aux communautés externes, sans tenir compte de l'intensité ou de la diversité de ces connexions, comme précisé dans la section \ref{subsec:nosmesures}. La classification ainsi obtenue ne sépare pas les capitalistes sociaux selon leurs degrés ou ratios. Ceux-ci se trouvent majoritairement dans les groupes de n\oe{}uds considérés comme périphériques et ultra-périphériques. 

Afin de dépasser les limites inhérentes à la participation, nous appliquons donc nos mesures proposées sur les données et présentons les résultats obtenus dans la sous-section qui suit.

\subsection{Groupes}
\label{proprietesgroupes}
Considérons tout d'abord les mesures obtenues sur l'ensemble des données traitées. On observe des corrélations positives pour l'ensemble des paires de mesures, allant de valeurs proches de $0$ à $0,9$. Les deux variantes d'une m\^eme mesure (liens entrants contre liens sortants) sont peu corrélées, ce qui confirme une nouvelle fois l'intér\^et de tenir compte de l'orientation dans notre étude. Trois mesures sont fortement corrélées : les intensités internes et externes et l'hétérogénéité ($\rho$ allant de $0,78$ à $0,92$). Le lien entre les intensités interne et externe semble indiquer que les variations dans le degré total d'un n\oe{}ud ont globalement le même effet sur ses degrés internes et externes. Autrement dit, la proportion entre ces deux types de liens ne dépend pas du degré du n\oe{}ud. Le très fort lien observé entre hétérogénéité et intensité indique que seuls les n\oe{}uds de faible intensité sont connectés de façon homogène à des communautés externes, tandis que les n\oe{}uds possédant de nombreux liens sont connectés de façon hétérogène.

En ce qui concerne l'analyse de regroupement, nous obtenons la meilleure séparation pour $k=6$ groupes, dont le Tableau \ref{tab:groupes} donne les tailles. Nous avons caractérisé les groupes relativement à nos huit mesures, afin d'en identifier les rôles et de les comparer à ceux définis par Guimerà et Amaral. Le Tableau \ref{tab:moyennes} contient les valeurs moyennes obtenues pour chaque mesure dans chaque groupe. Les {\sc ANOVA} que nous avons réalisées ont révélé des différences significatives pour toutes les mesures ($p<0.01$). Un test post-hoc ($t$-test avec correction de Bonferroni) a montré que ces différences existaient entre tous les groupes, pour toutes les mesures.

\begin{table}[h]
	\centering
	\begin{tabular}{|l|r|r|r|}
		\hline
		\textbf{Groupe} & \textbf{Taille} & \textbf{Proportion} & \textbf{Rôle} \\
		\hline
		1 & $24543667$ & $46,68\%$ & Non-pivot ultra-périphérique \\
		2 &      $304$ & $<0,01\%$ & Pivot orphelin \\
		3 &   $303674$ &  $0,58\%$ & Pivot connecteur	\\
		4 & $11929722$ & $22,69\%$ & Non-pivot périphérique (entrant) \\
		5 & $10828599$ & $20,59\%$ & Non-pivot périphérique (sortant) \\
		6 &  $4973717$ &  $9,46\%$ & Non-pivot connecteur \\
		\hline
	\end{tabular}
	\caption{Tailles de groupes détectés, et rôles correspondants dans la typologie de Guimerà et Amaral.}
	\label{tab:groupes}
\end{table}

Dans le groupe $1$, toutes les mesures sont négatives mais proches de $0$, à l'exception des deux variantes de la diversité, en particulier l'entrante, qui est proche de $-1$. Il ne peut pas s'agir de pivot au sens de Guimerà et Amaral (n\oe{}ud largement connecté  à sa communauté), puisque l'intensité interne est négative. De même, les mesures externes sont très faibles ce qui montre qu'il ne s'agit pas non plus de n\oe{}ud qualifiés de connecteurs par Guimerà et Amaral (ayant une connexion privilégiée avec d'autres communautés que la leur). On peut donc considérer que ce groupe correspond au role des non-pivots ultra-périphériques. Ce groupe est le plus grand (il contient à lui seul $47\%$ des n\oe{}uds), ce qui confirme la correspondance avec ce rôle, dont les n\oe{}uds constituent généralement la masse du réseau. Relativement au système modélisé, ces n\oe{}uds sont caractérisés par le fait qu'ils sont particulièrement peu suivis par les autres communautés.

Le groupe $4$ est extrêmement similaire au groupe $1$, à la différence que sa diversité entrante est de $0,69$. Ces n\oe{}uds restent donc périphériques, car l'intensité externe est toujours négative, mais ils reçoivent néanmoins des liens provenant d'un nombre relativement élevé de communautés. Autrement dit, ils sont suivis par peu d'utilisateurs externes, mais ceux-ci sont situés dans un grand nombres de communautés distinctes. Le groupe $5$ est lui aussi très proche du groupe $1$, mais la différence est cette fois que les deux variantes de la diversité sont positives, avec une diversité sortante de $0,60$. À l'inverse du groupe $4$, on peut donc dire ici que les utilisateurs concernés suivent (avec une faible intensité) des utilisateurs situés dans un grand nombre de communautés différentes. Les groupes $4$ et $5$ sont respectivement le deuxième ($23\%$) et troisième ($21\%$) plus grands groupes en termes de taille, ce qui porte le total des n\oe{}uds périphériques à $91\%$.

\begin{table}[h]
	\centering
	\begin{tabular}{|l|r|r|r|r|r|r|r|r|}
		\hline
		\textbf{G} &
 		\multicolumn{2}{|c|}{$\mathbf{I_{int}}$} & 
 		\multicolumn{2}{|c|}{$\mathbf{D}$} &
 		\multicolumn{2}{|c|}{$\mathbf{I_{ext}}$} &
 		\multicolumn{2}{|c|}{$\mathbf{H}$} \\
		\hline
		1 & $-0,12$ &  $-0,03$ & $-0,55$ & $-0,80$ &  $-0,09$ &  $-0,04$ &  $-0,12$ &  $-0,06$	\\
		2 & $94,22$ & $311,27$ &  $7,18$ & $88,40$ & $113,87$ & $283,79$ & $112,79$ & $285,57$	\\
		3 &  $5,52$ &   $1,40$ &  $5,60$ &  $3,10$ &   $5,28$ &   $1,43$ &   $6,76$ &   $2,34$	\\
		4 & $-0,04$ &   $0,00$ & $-0,37$ &  $0,69$ &  $-0,07$ &   $0,00$ &  $-0,10$ &  $-0,01$	\\
		5 & $-0,03$ &  $-0,01$ &  $0,60$ &  $0,19$ &  $-0,03$ &  $-0,02$ &  $-0,04$ &  $-0,02$	\\
		6 &  $0,48$ &   $0,12$ &  $1,96$ &  $1,70$ &   $0,35$ &   $0,12$ &   $0,53$ &   $0,19$	\\
		\hline
	\end{tabular}
	\caption{Mesures moyennes obtenues pour les $6$ groupes. Pour chaque mesure, deux valeurs sont indiquées, correspondant respectivement aux deux variantes : liens sortants et entrants.}
	\label{tab:moyennes}
\end{table}

Toutes les mesures sont positives dans le groupe $6$. L'intensité interne reste proche de $0$, donc on ne peut toujours pas parler de pivot, même si ces n\oe{}uds sont mieux connectés à leur communautés que ceux des groupes précédents. L'intensité externe est elle aussi faible, mais le fait qu'elle soit positive, à l'instar des autres mesures externes, semble suffisante pour considérer ces n\oe{}uds comme des connecteurs au sens de Guimerà et Amaral (relativement bien reliés à d'autres communautés). La diversité est relativement élevée, aussi bien pour les liens entrants que sortants ($D>1,7$). Ces n\oe{}uds sont donc plus fortement connectés à leur communauté mais aussi à l'extérieur, et avec une plus grande diversité. Il s'agit du quatrième plus gros groupe, représentant $9,5\%$ des n\oe{}uds.

Toutes les mesures du groupe $3$ sont largement positives : supérieures à $1,4$ pour celles basées sur les liens entrants, et supérieures à $5,2$ pour les liens sortants. L'intensité interne élevée permet d'associer ce groupe au rôle de pivot. Les valeurs externes montrent en plus que ces n\oe{}uds sont connectés à de nombreux n\oe{}uds présents dans de nombreuses autres communautés. Toutefois, les liens sortants sont plus nombreux, ces n\oe{}uds correspondent donc à des utilisateurs plus suiveurs que suivis. Ce groupe ne représente que $0,6\%$ des n\oe{}uds, il s'agit donc d'un rôle bien plus rare que ceux associés aux groupes précédents. Cette observation est encore plus caractéristique du groupe $2$, qui représente bien moins de $1\%$ des n\oe{}uds. Toutes les mesures y sont particulièrement élevées, la plupart dépassant $100$. Pour une mesure donnée, la variante concernant les liens entrants est toujours largement supérieure, ce qui signifie que les utilisateurs représentés par ces n\oe{}uds sont particulièrement suivis, et donc influents. Nous associons ce groupe au rôle de pivot orphelin defini par Guimerà et Amaral.

En conclusion de cette analyse des groupes, on peut constater que tous les rôles identifiés par Guimerà et Amaral ne sont pas présents dans le réseau étudié : on n'y trouve ni non-pivots orphelins, ni pivots provinciaux. Cette observation semble confirmer la nécessité d'une approche objective pour déterminer comment regrouper les n\oe{}uds en fonction des mesures. Elle est également consistante avec la forte corrélation observée entre les intensités interne et externe : les rôles manquants correspondraient à des n\oe{}uds possédant une forte intensité interne mais une faible intensité externe, ou vice-versa. Or, ceux-ci sont très peu fréquents dans notre réseau. De plus, le fait de distinguer les liens entrants et sortants permet d'obtenir une typologie plus fine. Ainsi, certains groupes distincts ont émergé (groupes $4$ et $5$) là où l'approche de Guimerà et Amaral aurait considéré ces n\oe{}uds comme équivalents.

\subsection{Positionnement des capitalistes sociaux}
\label{subsec:positionnement}
Avec la m\'ethode définie dans \cite{DUGUE2013}, nous d\'etectons près de $160.000$ capitalistes sociaux. Nous \'etudions ici leur positionnement dans les $6$ groupes identifi\'es par la m\'ethode des $k$-moyennes. De plus, nous affinons notre analyse en structurant les capitalistes sociaux en diff\'erents groupes. Tout d'abord via le ratio, qui nous permet de mettre en évidence les comportements {\sc FMIFY} et {\sc IFYFM}. Ensuite, en utilisant le degr\'e de ces utilisateurs. En effet, les capitalistes sociaux ayant accru le plus efficacement leur nombre de followers sont susceptibles d'avoir un placement ou un r\^ole diff\'erent au sein des communaut\'es.

Chaque tableau présente ainsi sur la première ligne la proportion de capitalistes sociaux du réseau qui sont contenus dans chaque groupe, et sur la deuxième la proportion de n\oe{}uds du groupe qui sont des capitalistes sociaux.\\

\noindent \textbf{Capitalistes sociaux de faible degré entrant.}

\begin{table}[h]
	\centering
	\begin{tabular}{|l|r|r|r|r|r|r|}
		\hline
		 \textbf{Ratio} & \textbf{G1}  & \textbf{G2} & \textbf{G3} & \textbf{G4} & \textbf{G5} & \textbf{G6}\\
		\hline
		  		$<1$ & $0.01\%$ & $0.00\%$ & $\mathbf{23.10\%}$ &  $3.42\%$ &  $\mathbf{18.28\%}$ & $\mathbf{55.19\%}$  \\
		   			 & $< 0.01\%$ & $0.00\%$ & $3.71\%$   &  $0.14\%$ &  $0.08\%$ & $0.54\%$  \\
		  \hline
		  		$>1$  & $0.03\%$   & $0.00\%$ & $\mathbf{18.78\%}$  &  $0.48\%$   &  $\mathbf{14.31\%}$ & $\mathbf{66.40\%}$ \\
		   			  & $< 0.01\%$ & $0.00\%$ & $\textbf{6.61\%}$ &  		   $< 0.01\%$ &   $0.14\%$ 				& $1.43\%$ \\
		\hline
	\end{tabular}
	\caption{Répartition des capitalistes sociaux de faible degré dans les différents groupes.}
	\label{tab:ksociaux500_generalized}
\end{table}

Ces capitalistes sociaux se retroupent dans trois groupes : $3$, $5$ et $6$. Les n\oe{}uds du groupe $3$ sont des pivots connecteurs qui ont en particulier tendance à suivre plus d'utilisateurs du réseau que la normale. M\^eme si le degré entrant des capitalistes sociaux est considéré comme faible ici, il reste élevé relativement au degré moyen du reste du réseau. Cela semble donc cohérent de voir qu'un grand nombre de capitalistes sociaux est plus connecté à la fois à leur communauté mais également aux autres communautés. Il semble également cohérent d'observer que les capitalistes sociaux de type \textbf{IFYFM} dont le degré sortant est supérieur au degré entrant sont près de deux fois plus présents dans ce groupe que les autres. La diversité sortante élevée du groupe $3$ nous apprend également que ces capitalistes sociaux ont tendance à ne pas cibler uniquement leur communauté m\^eme s'ils y sont bien connectés, mais à appliquer leurs méthodes à travers de nombreuses communautés du réseau.

On observe que la large majorité des capitalistes sociaux de faible degré se place au sein du groupe $6$, non-pivot connecteur. Ces n\oe{}uds, qui sont légèrement plus connectés au sein de leur communauté et avec l'extérieur que la moyenne, ont en revanche une diversité bien plus élevée. Les capitalistes sociaux qui s'y situent semblent ainsi avoir débuté l'application de leurs méthodes, en créant des liens avec de nombreuses autres communautés.

Enfin, on retrouve une faible proportion de capitalistes sociaux de faible degré dans le groupe $5$, groupe de n\oe{}uds non-pivots périphériques. Un certain nombre de capitalistes sociaux sont ainsi isolés au sein de leur communauté et avec l'extérieur.\\

\noindent \textbf{Capitalistes sociaux de degré entrant élevé.}

\begin{table}[h]
	\centering
	\begin{tabular}{|l|r|r|r|r|r|r|}
		\hline
		\textbf{Ratio} & \textbf{G1}  & \textbf{G2} & \textbf{G3} & \textbf{G4} & \textbf{G5} & \textbf{G6}\\
		\hline
		   $<0.7$ 	& $0.00\%$ & $\mathbf{12.14\%}$ & $\mathbf{87.29\%}$ & $0.00\%$  & $0.00\%$ & $0.57\%$ \\
		   			& $0.00\%$ & $\mathbf{21.05\%}$ & $0.15\%$ 	&  $0.00\%$ & $0.00\%$ & $< 0.01\%$		  \\
		   \hline
		    $>0.7$ et $<1$ & $0.00\%$ & $1.55\%$ 		  & $\mathbf{95.64\%}$ & $0.00\%$ & $0.00\%$ & $2.81\%$ \\
		   				   & $0.00\%$ & $\mathbf{7.24\%}$ & $0.45\%$  & $0.00\%$ &  $0.00\%$ & $< 0.01\%$		    \\
		   \hline
		    $>1$ & $0.00\%$ & $0.03\%$ & $\mathbf{97.99\%}$ & $0.00\%$ & $0.00\%$ & $1.98$ \\
		    	 & $0.00\%$ & $0.33\%$ & $1.22\%$ 	& $0.00\%$ & $0.00\%$ & $< 0.01\%$		 \\
		
		\hline
	\end{tabular}
	\caption{Répartition des capitalistes sociaux de degré élevé dans les différents groupes.}
	\label{tab:ksociaux10000_generalized}
\end{table}

Les capitalistes sociaux de degré élevé se placent presque exclusivement dans les groupes $2$ et $3$. Ces groupes contiennent des n\oe{}uds pivots connecteurs et orphelins.
Cela semble cohérent avec les degrés élevés de ces n\oe{}uds. Ceux-ci sont naturellement plus connectés avec leurs communautés et avec l'extérieur que les autres n\oe{}uds.
On constate que les n\oe{}uds classés dans le groupe $2$ sont ceux de ratio inférieur à $1$ et particulièrement ceux de ratio inférieur à $0,7$ ayant beaucoup plus de followers que de followees. Cela correspond bien à la définition du r\^ole donné par nos mesures qui montre que ce groupe de n\oe{}uds est suivi par un grand nombre de n\oe{}uds provenant d'une large variété de communautés.

En conclusion, on observe ainsi que notre approche permet d'établir une nette séparation entre capitalistes sociaux de faible degré, majoritairement connecteurs et non-pivots et ceux de degré élevé, classé comme pivots.
Par ailleurs, les r\^oles obtenus permettent également de discriminer les utilisateurs de ratios différents. Les capitalistes sociaux de degré élevé et de ratio inférieurs à $1$ sont par exemple les seuls à appartenir au groupe des pivots orphelins. Ce n'était pas le cas avec l'approche originale adaptée aux graphes orientés. Enfin, notre approche permet de mieux décrire les différents r\^oles obtenus gr\^ace aux trois mesures utilisées pour caractériser la connectivité du n\oe{}ud aux communautés auxquelles il n'appartient pas.

\section{Conclusion}
Dans cet article, notre but est de proposer une extension à la méthode définie par \cite{Guimera2005} pour caractériser le role communautaire de n\oe{}uds dans des réseaux complexes. Nous définissons d'abord une version orientée des mesures originales, puis nous les étendons pour qu'elle tiennent compte des différents aspects de la connectivité des n\oe{}uds (diversité, intensité et hétérogénéité). Nous proposons ensuite une méthode non-supervisée pour déterminer les rôles à partir de ces mesures. Elle a l'avantage d'être indépendante du système étudié. Enfin, nous donnons un exemple d'application en utilisant nos outils pour analyser le rôle des capitalistes sociaux dans Twitter. Notre méthode met en lumière les r\^oles caractéristiques joués par les capitalistes sociaux. Ceux de degré entrant élevé sont considérés comme des pivots orphelins ou connecteurs, en fonction de leur ratio. Ceux de faible degré entrant sont pour la plupart des non-pivots connecteurs. La prise en compte de l'orientation des liens, notamment, permet d'obtenir des r\^oles plus pertinents, ce qui confirme l'intérêt d'exploiter cette information lors de l'\'etude des r\'eseaux sociaux. 

Le travail présenté peut s'étendre de différentes façons. Tout d'abord, certains des rôles définis dans \citep{Guimera2005} n'apparaissent pas dans notre analyse. Il serait intéressant d'étudier d'autres réseaux afin de déterminer si cette observation reste valable. Une autre piste consiste à baser nos calculs sur des communautés recouvrantes (i.e. non-mutuellement exclusives). En effet, les réseaux sociaux que nous étudions sont réputés posséder ce type de structures, dans lesquelles un n\oe{}ud peut appartenir à plusieurs communautés en même temps \citep{Arora2012} ; de plus, de nombreux algorithmes existent pour les détecter \citep{Xie2013}. L'adaptation de nos mesures à ce contexte se ferait naturellement, en définissant des versions internes de l'hétérogénéité et de la diversité.
%TODO autre piste à laquelle je viens de penser : voir quels sont les noeuds-anomalies détectés à la suite du clustering. ceux qui ne rentrent pas vraiment dans un role.

\bibliographystyle{rnti}
\bibliography{dugue2014}

\appendix
%\section*{Annexe}
%
%Voici un exemple d'annexe. S'il y a plus de deux annexes, merci de
%les numéroter (Annexe~1, Annexe~2, etc).
%
\Fr

\end{document}